\documentclass[graybox,vecphys]{svmult}

\usepackage{makeidx}         
\usepackage{graphicx}        
\usepackage{multicol}        
\usepackage[bottom]{footmisc}

\usepackage{newtxtext}       %
\usepackage{newtxmath}       

\usepackage{url} 

\makeindex             

\begin{document}

\newcommand{\jcmindex}[2]{\index{{\bf\large #1}!#2}}
\newcommand{\jcmindext}[3]{\index{{\bf\large #1}!#2!#3}}

\title*{Nonlinear systems for unconventional computing}
\titlerunning{Nonlinear syst. for unconv. comp.}
\author{
Kirill P. Kalinin
\and
Natalia G. Berloff
}

\institute{
Kirill P. Kalinin
\at
University of Cambridge, Cambridge CB3 0WA, United Kingdom
\texttt{kpk26@cam.ac.uk}
\and
Natalia G. Berloff
\at
Skolkovo Institute of Science and Technology Russian Federation, Bolshoy Boulevard 30, bld. 1
Moscow, Russia 121205
\and
University of Cambridge, Cambridge CB3 0WA, United Kingdom
\email{N.G.Berloff@damtp.cam.ac.uk}
}
\maketitle
\abstract
{
The search for new \jcmindex{C}{computational} computational machines beyond the traditional von Neumann architecture has given rise to a modern area of \jcmindex{N}{nonlinear} nonlinear science -- development of \jcmindex{UC}{unconventional computing} unconventional computing -- requiring the efforts of mathematicians, physicists and engineers. Many \jcmindex{APS}{analogue physical systems} analogue physical systems including nonlinear oscillator networks, \jcmindex{L}{lasers} lasers, and \jcmindex{C}{condensate} condensates were proposed and realised to address \jcmindex{HCP}{hard computational problems} hard computational problems from various areas of social and physical sciences and technology. The \jcmindex{AS}{analogue systems} analogue systems emulate \jcmindex{SH}{spin Hamiltonian} spin Hamiltonians with continuous or discrete degrees of freedom to which actual \jcmindex{OP}{optimisation problems} optimisation problems can be mapped.  
Understanding of the underlying physical process by which the system finds the \jcmindex{GS}{ground state} ground state often leads to new classes of \jcmindex{SIA}{system-inspired algorithms} system-inspired or \jcmindex{QIA}{quantum-inspired algorithms} quantum-inspired algorithms for \jcmindex{HO}{hard optimisation} hard optimisation. 
Together physical platforms and related algorithms can be  combined to form a hybrid architecture that may one day compete with \jcmindex{CC}{conventional computing} conventional computing. In this Chapter, we review some of the systems and \jcmindex{PIA}{physically-inspired algorithms} physically-inspired algorithms that show such promise.}

\section{Introduction}
\label{sec:2}
We live in a world dominated by information. Systems that enable faster information processing and decision making are becoming more integrated into our daily lives. This data-intensive science relies on continual improvements in hardware for solving ever growing, e.g. in number of variables and constraints, optimisation problems. Digital electronics can no longer satisfy this trend, as exponential hardware scaling (Moore's law)  and the von Neumann architecture are reaching their limits \cite{MoorLawPaper1965,MoorLawOver2016}. 

Looking \jcmindex{BTC}{beyond traditional computing} beyond the traditional computing one turns to physical platforms that with their superior speed and reconfigurability and internal parallel processing can provide faster alternatives to solving a specialised class of \jcmindex{N}{nonlinear} nonlinear problems.  Despite a number of physical systems that were proposed as quantum or \jcmindex{AS}{analogue simulators} analogue simulators and further elucidated in active applied research, significant challenges still remain before scalable analogue processors can be realised and show the superior performance in comparison with the von Neumann computing architecture.  
Over the years, various \jcmindex{UC}{unconventional computing} unconventional computing techniques were proposed that enable  simultaneous communication, computation, and memory access throughout their  architecture with the purpose to alleviate the device and system architectural challenges faced by conventional computing platforms. 

\jcmindex{NC}{neuromorphic computing} Neuromorphic computing based on neural networks promises to make processors that use low energies while integrating massive amounts of information. \jcmindex{QAD}{quantum annealer devices} Quantum annealer devices promise to find the \jcmindex{GM}{global minimum} global minimum of a combinatorial \jcmindex{O}{optimisation} optimisation problem faster than classical computers. Physical (natural) systems aim to become \jcmindex{AM}{analogue machines} analogue machines  by bridging the physics of a particular system with engineering platforms enhancing performance of machine learning.

A central challenge is in the development of mathematical models -- \jcmindex{SIC}{system-inspired computing} system-inspired computing -- linking  physical platforms to models of complex analogue information processing.  Among such models, those based on principles of neural networks and quantum annealing are perhaps the most widely studied.

A large class of problems that can be solved on physical platforms includes  \jcmindex{N}{nonlinear} nonlinear programming problems. They  seek to minimise some \jcmindex{N}{nonlinear} nonlinear  objective function $E({\bf x})$ of real or complex variables in ${\bf x}$ subject to a series of constraints represented by equalities or inequalities, i.e.,  $g({\bf x}) = 0$ and $h({\bf x})\le 0$. Numerous applications in social sciences and telecommunications, finance and aerospace,  biological and chemical industries can be described in this basic framework \cite{applications,applications2,applications3}.
 
\jcmindex{N}{Nonlinear} Nonlinear  \jcmindex{O}{optimisation} optimisation problems are notoriously difficult to solve, and often involve specialised techniques such as genetic algorithms, particle swarm \jcmindex{O}{optimisation} optimisation, simulation and population annealing. Around the  vicinity of the \jcmindex{OS}{optimal solution} optimal solution \jcmindex{N}{nonlinear} nonlinear  \jcmindex{O}{optimisation} optimisation problems are  quadratic to second order, and therefore, \jcmindex{QP}{quadratic programming} Quadratic Programming (QP) for minimising  quadratic functions of variables subject to linear constraints is a usual simplification to such problems that can be used with a wide array of applications. QP occurs in various machine learning problems, such as the support vector machine (SVM) training and least squares regression. At the same time, QP and other \jcmindex{N}{nonlinear} nonlinear  \jcmindex{O}{optimisation} optimisation problems can be mapped to \jcmindex{SH}{spin Hamiltonians} spin Hamiltonians which can be emulated by real physical systems: the degrees of freedom ${\bf x}$  become `spins,' the cost function $E({\bf x})$  is a `Hamiltonian' that specifies the interaction pattern between spins. In this chapter we discuss two possible ways by which the system can find the optimal solution -- the \jcmindex{GS}{ground state} ground state of the corresponding \jcmindex{SH}{spin Hamiltonian} spin Hamiltonian -- depending on the nature of the system. The system in thermodynamic equilibrium may find the \jcmindex{OS}{optimal solution} optimal solution by  quantum annealing which is executed with the time-dependent Hamiltonian 
 \begin{equation}
 H(t)=\biggl(1- \frac{t}{\tau}\biggr) H_0 + \frac{t}{\tau} H_{\rm objective},
 \label{qa}
 \end{equation}
 where $H_0$ is the initial trivial Hamiltonian whose ground
state is known, and $H_{\rm objective}$ is the final Hamiltonian at $t=\tau$ which
encodes an original objective function $E({\bf x})$. If
the system is in thermal equilibrium at all times then it stays close to the
ground state, as Hamiltonian parameters are adiabatically varied.  A linear time dependence in Eq. (\ref{qa}) is assumed for simplicity but more complex annealing schedules can be used.  The time $\tau$ for obtaining the result of \jcmindex{O}{optimisation} optimisation is much larger than that defined by the inverse of the spectral gap (the distance between the ground state and the lowest excited state) of $H(t)$ \cite{farhi}. When spectral gap is large,  the coupling to the environment  helps
the annealer by  cooling the system towards its ground state, however, as the system becomes larger and the spectral gap shrinks (typically exponentially  fast with the system size) the excited states lead to large errors at the same time slowing down the annealing procedure. 

\jcmindex{NES}{non-equilibrium systems} Non-equilibrium systems rely on a different principle of approaching the \jcmindex{GS}{ground state} ground state from below rather than via quantum tunnelling during the adiabatic annealing. 
The principle of the \jcmindex{GDS}{gain-dissipative simulator} gain-dissipative simulator is based on  a two-stage process:  gain increase below the threshold and the coherence of operations at  the threshold. Ramping up the gain allows system to overcome its linear losses and to stabilise by the \jcmindex{N}{nonlinearity} nonlinearity of the  gain saturation. The emergent coherent state minimises the losses and, therefore, maximises the total number of particles as it will be explained further below, which leads to minimising a particular  functional that can be written as the objective spin Hamiltonian. Close to the threshold, the resulting evolution of the system elements resembles the dynamics of \jcmindex{HN}{Hopfield networks} Hopfield networks which were shown to be able to solve quadratic \jcmindex{O}{optimisation} optimisation problems more than thirty years ago \cite{Hopfield1985, QuestioningHopfield1988} by using  a system of differential equations that describe the evolution of individual neurons:
\begin{equation}
	 \frac{d x_i}{dt} = -\sum_{j=1}^N J_{ij} S_j(x_j),\label{hopfield} 
	 \end{equation} 
where $x_i$ is an input, $S_j(x_j) $ is the activation function (e.g. sigmoid), and $J_{ij}$ is the connectivity matrix among the neurons.
Various modifications of Hopfield networks  were extensively proposed and studied \cite{HopfieldModification1990}, however,  the optimisers based on  \jcmindex{HN}{Hopfield networks} Hopfield networks were surpassed by other computational methods.  This is largely due to the high connectivity between neurons that neural networks require and the concomitant time it takes to evolve large networks on classical hardware. The recent interest in \jcmindex{HN}{Hopfield networks} Hopfield networks re-emerged as it became possible to create them in \jcmindex{APS}{analogue physical systems} analogue physical systems such as electronic circuits  or photonic neural networks. Photonic systems have an advantage over their electronic counterparts due to the picosecond to femtosecond time scale of their operation and as  hundreds of high bandwidth signals can  flow through a single optical waveguide.  This means that a photonic implementation of \jcmindex{HN}{hopfield networks} Hopfield networks as optimisers  can  have a large dimensionality and dense connectivity as well as a fast convergence time. However, the evolution of \jcmindex{HN}{Hopfield networks} Hopfield networks does not necessarily lead to the \jcmindex{OS}{optimal solution} optimal solution.

\begin{figure}
\centering
\includegraphics[width=.9\textwidth]{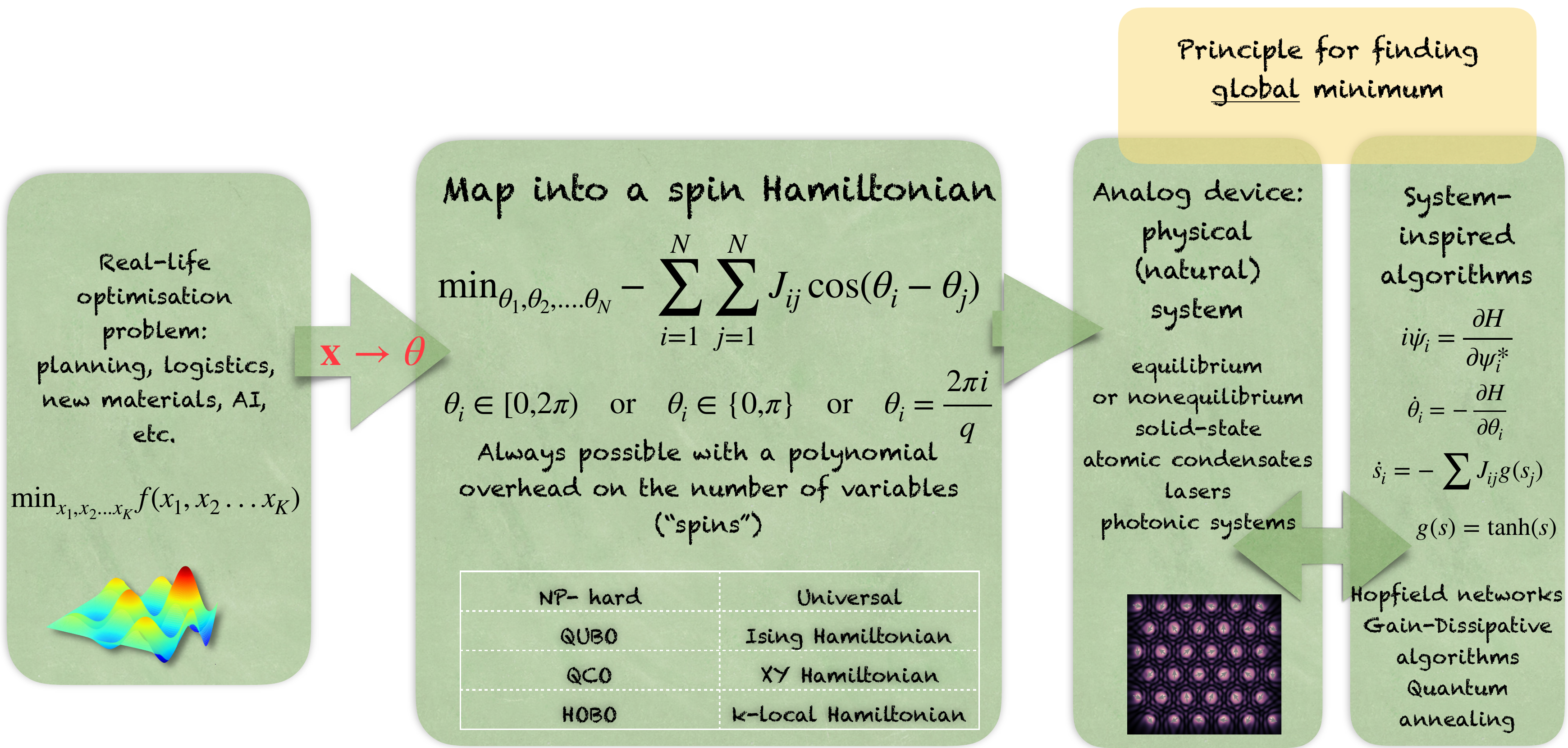}
\caption{Schematics of \jcmindex{HC}{hybrid computing} Hybrid Computing.  A computationally hard real life problem can be mapped into a \jcmindex{SH}{spin Hamiltonian} spin Hamiltonian, where spins represent the degrees of freedom (discrete or continuous) of the problem and coupling strengths represent the structure and the constraints of the objective function. Such mapping is possible if the original problem is nondeterministic polynomial, and the resulting \jcmindex{SH}{spin Hamiltonian} spin Hamiltonian is universal. The \jcmindex{GS}{ground state} ground state of the \jcmindex{SH}{spin Hamiltonian} spin Hamiltonian can be found by an \jcmindex{APS}{analogue physical system} analogue physical system using quantum annealing or the \jcmindex{GD}{gain-dissipative} gain-dissipative principle of its operation coupled to a digital device (CMOS, FPGA, GPA, etc.) compatible and based on the \jcmindex{SIA}{system-inspired algorithms} system-inspired algorithms.}
\label{schematicsComputing}       
\end{figure}

In this Chapter we will review  recent progress in building the \jcmindex{AD}{analogue devices} analogue devices that implement either quantum annealing or \jcmindex{GD}{gain-dissipative} gain-dissipative principle in their architecture while illustrating the idea of devising \jcmindex{SIA}{system-inspired algorithms} system-inspired algorithms.

\subsection{Spin Hamiltonians}
\label{subsec:1}

The majority of optimisation problems \jcmindex{OP}{optimisation problems} are computationally impractical for conventional classical computers with classic examples of a so-called \jcmindex{HO}{hard optimisation} ``hard optimisation task" being the \jcmindex{TSP}{travelling salesman problem} travelling salesman problem, the dynamic analysis of financial markets, the prediction of new chemical materials, and \jcmindex{ML}{machine learning} machine learning \cite{NN_npcomplete1989}. Mathematically, it is possible to reformulate many of these optimisation problems from vastly different areas as a problem of finding the \jcmindex{GS}{ground state} ground state of a particular \jcmindex{SH}{spin Hamiltonian} spin Hamiltonian with discrete or continuous degrees of freedom, as we will refer to, throughout this Chapter, simply as solving spin model. The \jcmindex{SH}{spin Hamiltonian} spin Hamiltonian can be  emulated with a given simulator, e.g. solid-state system, that would need to have an easy mapping of the variables of the desired Hamiltonian into the elements (spins, currents etc.) of the simulator, independently tunable short and long range interactions between them, and would allow one to perform measurements to obtain the answer with the required precision. Such \jcmindex{SH}{spin Hamiltonian} spin model Hamiltonians are experimentally challenging to implement and control but their possible advantageous performance over classical computers, which struggle solving sufficiently large problem sizes, leads to an intensive search for a superior simulator. Such simulators have been proposed and realised to a various extent in disparate physical systems.  Among these systems, two classes of \jcmindex{SH}{spin Hamiltonian} spin Hamiltonians are more common: \jcmindex{IH}{Ising Hamiltonian} \jcmindex{I}{Ising} Ising and \jcmindex{XY}{XY Hamiltonian} XY Hamiltonians. For instance, the \jcmindex{IH}{Ising Hamiltonian} Ising Hamiltonian  is  widely used for  a vast variety of hard discrete combinatorial \jcmindex{O}{optimisation} optimisation problems, so that travelling salesman, graph colouring, graph partitioning, and others can be mapped into it with a polynomial overhead \cite{Lucas2014}. This model is formulated for $N$ classical ``spins" $s_j$ that take discrete values $\{-1,1\}$ to minimise the \jcmindex{QUBO}{quadratic unconstrained binary optimisation} quadratic unconstrained binary optimisation (QUBO) problem: 
\begin{equation}
\min \quad -\,\sum_{i=1}^N \sum_{j=1,j<i}^N J_{ij} s_i s_j + \sum_{i=1}^N h_i s_i \quad {\rm subject} \ {\rm to} \quad s_i \in \{-1,1\}
\label{ising}
\end{equation}
where $h_i$ represents external (magnetic) field. This term can be incorporated in $\mathbb{J}$ matrix by considering $N+1$ spins and thus will be omitted for the rest of the chapter.
Experimental realisation of the \jcmindex{N}{nonlinear} nonlinear  terms beyond quadratic in the \jcmindex{IH}{Ising Hamiltonian} Ising Hamiltonian would lead to a $k$-local \jcmindex{SH}{spin Hamiltonian} spin Hamiltonian with $k > 2$ and would allow for a direct mapping of \jcmindex{HOBO}{higher order binary optimisation} higher order binary optimisation (HOBO) problems  including Max-SAT \cite{maxsat} or  number factorisation \cite{FactorisationMicrosoft}:

\begin{equation}
\min \quad -\,\sum_{i_1,i_2,...i_k}^N Q_{i_1,i_2,...i_l,...,i_k} s_{i_1} s_{i_2}...s_{i_l}...s_{i_k} \quad {\rm subject} \ {\rm to} \quad s_{i_l} \in \{-1,1\}.
\label{hobo}
\end{equation}
In the \jcmindex{XY}{XY Hamiltonian} XY model ``spins" are continuous $s_j = \cos \theta_j + i \sin \theta_j$ and the corresponding \jcmindex{QCO}{quadratic continuous optimisation} quadratic continuous optimisation (QCO) problem can be formulated as
\begin{equation}
\min -\,\sum_{i<j} J_{ij} {\bf s}_i \cdot {\bf s}_j = \min -\sum_{i<j} J_{ij} \cos(\theta_i - \theta_j) \quad {\rm subject} \ {\rm to} \quad \theta_i \in [0,2\pi).
\label{xy}
\end{equation}
When possible phases $\theta_j$ are limited to discrete values $2\pi /n$ with an integer $n>2$ the model (\ref{xy}) recovers the $n-$state \jcmindex{PH}{Potts Hamiltonian} Potts model (Clock model) with applications in protein folding \cite{PottsProteins}.  

\jcmindex{QCO}{quadratic continuous optimisation} QCO, \jcmindex{QUBO}{quadratic unconstrained binary optimisation} QUBO, and \jcmindex{HOBO}{higher order binary optimisation} HOBO problems are all examples of \jcmindex{NPH}{NP-hard} $\mathbb{NP}$-hard problems. The corresponding spin models are universal. The connection between these notions are detailed in the next section. 
\subsection{P, NP, NP-complete problems}
\label{subsec:2}
The \jcmindex{CS}{computational complexity} computational complexity of a problem can be revealed by looking at the dependence of the problem's size on time or the number of operations required to solve it. In a simple case of such polynomial dependence, i.e. when a polynomial time algorithm exists, a problem belongs to a $\mathbb{P}$ class. If a polynomial time algorithm of finding a solution is not known but there exists a polynomial algorithm for verifying a solution when presented, then a problem belongs to non-deterministic polynomial-time ($\mathbb{NP}$) class that clearly includes the $\mathbb{P}$ class. Whether $\mathbb{P}$ = $\mathbb{NP}$ is true or not is a major unsolved problem in computer science although it is widely believed to be untrue \cite{PvsNP_1992}. A problem is \jcmindex{NPH}{NP-hard} $\mathbb{NP}$-hard when every problem in $\mathbb{NP}$ can be reduced in polynomial time to it. The problems that are both \jcmindex{NPH}{NP-hard} $\mathbb{NP}$-hard and $\mathbb{NP}$ are called $\mathbb{NP}$-complete. All \jcmindex{NPC}{NP-complete} $\mathbb{NP}$-complete problems are equivalent in a sense that either all of them or none of them admit a polynomial-time algorithm. Examples include the \jcmindex{TSP}{travelling salesman problem} travelling salesman problem, spin glass models, and integer linear programming. 
The \jcmindex{CS}{computational complexity} computational complexity of finding the \jcmindex{GS}{ground state} ground state of the \jcmindex{IH}{Ising Hamiltonian} Ising Hamiltonian (Ising model) on finite lattices has been studied before \cite{Barahona1982} where the two-dimensional \jcmindex{IH}{Ising Hamiltonian}  Ising model with a magnetic field (\ref{ising}) and equal antiferromagnetic couplings  has been shown to be \jcmindex{NPC}{NP-complete} $\mathbb{NP}$-complete for planar graphs. In addition, $\mathbb{NP}$-completeness was demonstrated for the three-dimensional \jcmindex{IH}{Ising Hamiltonian} Ising model with nearest neighbour interactions and coupling strengths from $\{- 1, 0, 1\}$ \cite{Barahona1982}. Consequently, the above mentioned hierarchy of complexity classes allows one to conclude the impossibility of existence of a polynomial algorithm for computing the ground state energy of the \jcmindex{IH}{Ising Hamiltonian} Ising model without the existence of a polynomial algorithm for all \jcmindex{NPC}{NP-complete} $\mathbb{NP}$-complete problems.

The existence of universal \jcmindex{SH}{spin Hamiltonians} spin Hamiltonians has been established. Universality means that all classical spin models with any range of interactions can be reproduced within such a model, and certain simple Hamiltonians such as the 2D \jcmindex{IH}{Ising Hamiltonian} Ising model on a square lattice with transverse fields and nearest neighbour interactions are universal \cite{Cubitt_universality}. Thus, due to \jcmindex{NPH}{NP-hard} $\mathbb{NP}$-hardness of the \jcmindex{IH}{Ising Hamiltonian} Ising model, there should exist a polynomial time mapping of many practically relevant \jcmindex{NPC}{NP-complete} $\mathbb{NP}$-complete problems to the \jcmindex{IH}{Ising Hamiltonian} Ising Hamiltonian, whose decision version solves the \jcmindex{NPC}{NP-complete} $\mathbb{NP}$-complete problem of interest. The mapping of various $\mathbb{NP}$ problems, including Karp's 21 \jcmindex{NPC}{NP-complete} $\mathbb{NP}$-complete problems \cite{karp1972reducibility}, to \jcmindex{IH}{Ising Hamiltonian} Ising models with a polynomial overhead was demonstrated \cite{Lucas2014}. For example, the \jcmindex{TSP}{travelling salesman problem} travelling salesman problem for $N$ cities, that are connected with weighted edges $w_{uv} \ge 0$ from the set $E$ (distances between cities), can be formulated as the following Ising problem of size $N^2$:
\begin{eqnarray}
H_{\rm TSP} &=& A\sum_{i=1}^{N} \biggl(1 - \sum_{v=1}^N x_{v,i}\biggr)^2 +  A\sum_{v=1}^{N} \biggl(1 - \sum_{i=1}^N x_{v,i}\biggr)^2+ A\sum_{(uv)\notin E} \sum_{i=1}^N x_{u,i}x_{v,i+1} \nonumber \\
&+&B\sum_{(uv)\in E}w_{u,v} \sum_{i=1}^N x_{u,i}x_{v,i+1}.
\label{tsp1}
\end{eqnarray}
Each spin $x_{v,i} \in \{0,1\}$  in Eq.~(\ref{tsp1}) represents the vertex $v$ and its order $i$ in a path. All valid routes in this representation are regulated by the first three terms: each city should be in the route (first term) and should appear in it only once (second term), any adjacent cities in the route should be connected (third term), while the search for the optimal route is realised by minimising  the sum of weights of all cities in a route (forth term). The reasonable choice of constants $A$ and $B$ (e.g. $A$ should be big enough with respect to $B>0$) guarantees that only the space of valid routes is explored. Reshaping this two-dimensional spin matrix with elements $x_{v,i}$ to a spin vector of size $N^2$ allows one to recover the coupling matrix $\mathbb{J}$ and magnetic field ${\bf h}$ to formulate the corresponding \jcmindex{IH}{Ising Hamiltonian} Ising Hamiltonian. The size of the \jcmindex{IH}{Ising Hamiltonian} Ising problem can be reduced to $(N-1)^2$ by fixing a particular city to be the first in the route. Note, that the Hamiltonian $H_{\rm TSP}$ can represent both directed and undirected graphs, and the generalisation for the cycles optimisation problem is straightforward. We also note that a polynomial overhead does not always apply and some combinatorial optimisation problems can be mapped to the \jcmindex{IH}{Ising Hamiltonian} Ising model of the same size $N$. For example, the \jcmindex{MC}{maximum cut} maximum cut (MaxCut) problem
\begin{equation}
	\max_{S^+,S^-} \sum_{i \in S^+,j\in S^-} w_{ij}
\end{equation}
seeks for the cut of a graph into two subsets with a largest sum of their connecting weighted edges. By assigning $+1$ and $-1$ spins to all vertices in subsets $S^+$ and $S^-$, respectively, this optimisation problem can be formulated as
\begin{equation}
	\max_{s_i} \frac{1}{2} \sum_{i<j} w_{ij} (1 - s_i s_j) =  \frac{1}{2} \sum_{i<j} w_{ij} + \min_{s_i} \frac{1}{2} \sum_{i<j} w_{ij} s_i s_j
\end{equation}
and thus a maximum cut of any graph can be converted to minimisation of the corresponding \jcmindex{IH}{Ising Hamiltonian} Ising Hamiltonian with the \jcmindex{CM}{coupling matrix} coupling matrix $J_{ij} = - w_{ij}$ with an addition of an offset. A well-known standardised set of \jcmindex{MC}{maximum cut} MaxCut type of problems often serve as a metric for comparison of newly proposed simulators and algorithms \cite{BLS2013,GainD_sciRep_2018,CIM_eqs2019}.

Another example of a universal spin model is the \jcmindex{XY}{XY Hamiltonian} XY model which is directly related to the notoriously hard to solve phase retrieval problem. The problem's objective is to recover a general signal (or image) from the magnitude of its Fourier transform \cite{PhaseRetrieval1,PhaseRetrieval2,PhaseRetrieval3}. This problem arises from the fact that the signal detectors can usually record only modulus of the diffraction pattern, therefore, losing the information about the phase of the optical wave. Mathematically, one needs to recover a signal ${\bf x} \in \mathbb{C}^m$ from the amplitude ${\bf b} = | A {\bf x} |$, where $A \in \mathbb{C}^{n \times m}$, ${\bf b} \in \mathbb{R}^n$. Then the phase recovery problem \cite{PhaseRetrievalToXY} can be formulated as:
\begin{equation}
	\min_{x_j, u_i} \sum_i \bigg( \sum_j A_{ij} x_j - b_i u_i \bigg)^2
\end{equation}
where  ${\bf u} \in \mathbb{C}^n$ is a phase vector that satisfies $A {\bf x} = diag( {\bf b}) {\bf u}$, $|u_i| = 1$ for $i=\overline{1,n}$. This optimisation problem can be further rewritten as
\begin{equation}
\min \sum_{ij} M_{ij} u_i u_j \quad {\rm subject} \ {\rm to} \quad |u_i| = 1, i = \overline{1,n},
\end{equation}
where $M = diag({\bf b}) (I - A A^\dagger) diag({\bf b})$ is the Hermitian matrix, $I$ is the identity matrix, and $A^\dagger$ is the Moore-Penrose pseudoinverse of a matrix $A$ (see \cite{PhaseRetrievalToXY} for details). 

It is important to note that when we refer to a spin problem as \jcmindex{NPC}{NP-complete} $\mathbb{NP}$-complete we understand that for some specific coupling matrix $\mathbb{J}$ ('problem instances')  finding the solution can be easy (belong to $\mathbb{P}$ class).  The term \jcmindex{NPC}{NP-complete} $\mathbb{NP}$-completeness reflects worst case behaviour and may allow a polynomial time to solution for most instances on average. This leads to the cornerstone question of how to distinguish hard instances from simple ones. The answer is especially important for the rivalry between classical hardware and \jcmindex{UC}{unconventional computing} unconventional computing machines which have to compete on problems of known complexity. It is believed that the way to create ``hard" instances for \jcmindex{SH}{spin Hamiltonian} spin Hamiltonians resides at the intersection of computational complexity and physics, e.g. the hardness of problems can be connected to the existence of a first-order phase transition in a system (see \cite{Helmut_wishart} and references therein). If an instance is indeed hard then it would be difficult to solve even for a medium size on a classical computer since the number of operations grows as an exponential function with the matrix size. Thus, the time required to find reliably the ground state energy should highly depend on the \jcmindex{CM}{coupling matrix} coupling matrix structure $\mathbb{J}$ and the way it was constructed. For instance, finding the global minimum of the \jcmindex{XY}{XY Hamiltonian} XY model for positive definite matrices remains \jcmindex{NPH}{NP-hard} $\mathbb{NP}$-hard due to the non-convex constraints but can be effectively approximated using a semidefinite programming (SDP) relaxation with some performance guarantee \cite{SDP_programming1,SDP_programming2}. Sparsity also plays an important role and for sufficiently sparse matrices fast methods exist \cite{SDP_programming3}. For spin models, the generation of matrix instances $\mathbb{J}$ with tunable algorithmic hardness and, preferably, with a specifiable ground state, is an ongoing problem studied by many research teams. An elegant way of creating such problems has been recently proposed by a Microsoft research team \cite{Helmut_wishart} who suggested to use the Wishart planted ensemble technique. Having a unified set of optimisation problems with a tunable hardness and known solutions would allow for an objective benchmark of quantum simulators on various physical platforms as well as for classical algorithms. Otherwise, announcements of state-of-the-art platforms and methods, which demonstrated their performance on some random and not necessarily hard instances, would continue to happen.

\section{Physical platforms for large-scale  optimisation}
\label{sec:2}
Rather than trying to model nature one can consider a reverse idea of exploiting physical phenomena for solving \jcmindex{NPC}{NP-complete} $\mathbb{NP}$-complete problems. Such problems can be tackled by \jcmindex{QC}{quantum computers} quantum computers or \jcmindex{QS}{quantum simulators} simulators to produce solutions in reasonable time. In the last five years we have seen a competition of different physical platforms in solving classical optimisation problems faster than it can be achieved on a classical hardware for a given problem size. This rivalry resulted in the rapid emergence of a new field at the intersection of \jcmindex{L}{laser}laser and condensed matter physics, engineering and complexity theories, which aims to develop \jcmindex{QAD}{quantum analogue devices} quantum or \jcmindex{CAD}{classical analogue devices} classical analogue devices to simulate \jcmindex{SH}{spin Hamiltonians} spin Hamiltonians. Next we discuss the achieved success in such simulations for a range of physical systems.


\subsection{Cold atoms in optical lattices}
\label{subsec:1}
\jcmindex{UA}{ultracold atoms} Ultracold atoms in \jcmindex{OL}{optical lattice} optical lattices constitute a well-controlled experimental setting to realise various \jcmindex{SH}{spin Hamiltonians} spin Hamiltonians \cite{ultracoldBEC1,ultracoldBEC2}. \jcmindex{OL}{optical lattice} Optical lattices are formed by directing several \jcmindex{L}{laser}laser beams to interfere and to create standing wave configurations. Such waves provide practically loss-free external potentials in which \jcmindex{UA}{ultracold atoms} ultracold atoms may condense, move and
interact with one another \cite{grimm2000,windpassinger2013engineering}. The unprecedented control and precision with which one can engineer such lattices and load the atoms there led to many suggestions to
consider such systems as possible candidates for \jcmindex{UC}{unconventional computing} unconventional computing in quantum information processing and \jcmindex{QS}{quantum simulations}  quantum simulations. 

Here we will only discuss  a weakly interacting \jcmindex{B}{Bose}  Bose gas in an \jcmindex{OL}{optical lattice}  optical lattice. The description of particles in the strongly-correlated regime  is possible with \jcmindex{B}{Bose}  Bose- and Fermi-Hubbard models as well as with  extended Hubbard models with \jcmindex{NN}{nearest-neighbour}  nearest-neighbour, next nearest-neighbour  interactions, etc. \cite{hubbard}.
If the bosonic gas is dilute, the time evolution of the \jcmindex{C}{condensate}condensate wave function $\psi$  is governed by the \jcmindex{GPE}{Gross-Pitaevskii equaiton}  Gross-Pitaevskii equation (GPE) \cite{gpe,gpe2,pitaevskii2016bose}
\begin{equation}
i\hbar \frac{d}{dt}\psi({\bf r},t)=-\frac{\hbar^2}{2m}\nabla^2\psi({\bf r},t)+V_{\rm ext}({\bf r})\psi({\bf r},t)+g|\psi({\bf r},t)|^2\psi({\bf r},t),
\label{gpe}
\end{equation}
where $g$ is the strength of the delta-function interactions and the external potential $V_{\rm ext}$ describes an optical lattice -- periodic potential -- usually combined with a weak  harmonic trapping potential.

The \jcmindex{C}{condensate}condensate evolution and particles' interactions at different local minima of the \jcmindex{OL}{optical lattice} optical lattice in superfluid regime can be described with the tight-binding approximation, which is valid when the barrier between the neighbouring sites is much higher than the chemical potential. 
In this approximation the \jcmindex{C}{condensate}condensate wavefunction  $\psi$ is written as  a
sum of normalized wave functions $\phi_i=\phi({\bf r}-{\bf r}_i)$ localized in each
minimum of the periodic potential, i.e. ${\bf r}={\bf r}_i$:
\begin{equation}
\psi({\bf r},t)=\sum_i \Psi_i(t)\phi({\bf r}-{\bf r}_i),
\label{ansatz}
\end{equation}
where $\Psi_i(t)=\sqrt{\rho_i(t)}e^{i\theta_i(t)}$ is the complex amplitude of
the $i$--th lattice site,  $\rho_i$ and $\theta_i$ are the number of particles and the phase 
 in the $i$--th site, respectively. The amplitude  $\Psi_i$  describes the state of the so-called `coherent center' located at ${\bf r}_i$. By inserting this anzats into Eq.~(\ref{gpe}) and integrating the spatial degrees of freedom out   
one obtains the discrete \jcmindex{N}{nonlinear} nonlinear  Schr\"odinger (DNLS) equation  (see e.g. \cite{eilbeck1985discrete,PhysRevE.66.046608,Trombettoni01} 
\begin{equation}
i\hbar\frac{\partial\Psi_i}{\partial t}=-J(\Psi_{i+1}+\Psi_{i-1})+\epsilon_i\Psi_i+U|\Psi_i|^2\Psi_i,\,
\label{dnlse}
\end{equation}
 where $J$ is the \jcmindex{NN}{nearest-neighbour} nearest-neighbour tunnelling rate, 
 \begin{equation}
J=-\int d{\bf r}\left[\frac{\hbar^2}{2m}{\bf \nabla}\phi_i\cdot{\bf \nabla}\phi_{i+1}+\phi_i
V_{\rm ext}\phi_{i+1}\right], \label{tunneling}
\end{equation}
 $\epsilon_i$ is 
the on-site energy given by
\begin{equation}
\epsilon_i=\int d{\bf r}\left[\frac{\hbar^2}{2m}({\bf \nabla}\phi_i)^2+V_{\rm ext}\phi^2_{i}\right],
\label{onsite}
\end{equation}
and $U$ is the  \jcmindex{N}{nonlinear} nonlinear coefficient given by
\begin{equation}
U=g\int d{\bf r}\,\phi^4_{i}\,.
\label{nonlinear}
\end{equation}
Such classical lattice models described by DNLS equations represent the mean-field limit of Bose--Hubbard models \cite{Mishmash2009}. The \jcmindex{MF}{mean-field}  mean-field limit of the non-standard Bose-Hubbard models includes  the interactions beyond the nearest neighbours which leads to a generalised DNLS
\begin{equation}
i\hbar\frac{\partial\Psi_i}{\partial t}=-\frac{1}{2}\sum_{\langle i,j\rangle}J_{ij}\Psi_j+(\epsilon_i+U|\Psi_i|^2)\Psi_i,\,
\label{dnlse2}
\end{equation}
where $J_{ij}$ is the coupling strength between the $i$-th and $j$-th coherent centers. 
If one loads an equal number of particles in each site of the lattice, 
the \jcmindex{GS}{ground state} ground state of Eq.~(\ref{dnlse2}) realises the minimum of the \jcmindex{XY}{XY Hamiltonian} XY Hamiltonian $-\sum_{\langle i,j\rangle}  J_{ij} \cos(\theta_i-\theta_j)$. This has been experimentally demonstrated in  triangular lattices using the atoms motional degrees of freedom  and tunable artificial gauge fields \cite{struck11, struck13}.

The quantum annealing protocol can in principle be implemented in such a system  by  using  Eq.~(\ref{qa}) with 
$H_0=\sum_{\langle i,j\rangle} \cos(\theta_i-\theta_j)$ and $H_{\rm objective}=-\sum_{\langle i,j\rangle} J_{ij} \cos(\theta_i-\theta_j).$ 
A similar principle of adiabatic quantum annealing has been realised in the famous \jcmindex{DW}{D-Wave} D-Wave machine that we discuss below.

\subsection{D-Wave quantum annealer}
\label{subsec:5}

\jcmindex{DW}{D-Wave} D-Wave is a first commercially available quantum annealer that is built on superconducting qubits with programmable couplings and specifically designed to solve \jcmindex{QUBO}{quadratic unconstrained binary optimisation} QUBO problems (\ref{ising}) \cite{Dwave1}. By specifying the interactions $J_{ij}$ between qubits, a desired \jcmindex{QUBO}{quadratic unconstrained binary optimisation} QUBO problem is solved  \cite{Dwave2} via a quantum annealing process as in Eq.~(\ref{qa}). Adiabatic (slow) transition in time from an initial state of a specially prepared ``easy" Hamiltonian to the objective \jcmindex{IH}{Ising Hamiltonian} Ising Hamiltonian guarantees that the system remains in the low energy state, which gives the final energy that corresponds to the \jcmindex{OS}{optimal solution} optimal solution of the \jcmindex{QUBO}{quadratic unconstrained binary optimisation} QUBO problem.

Many benchmarks on different \jcmindex{QUBO}{quadratic unconstrained binary optimisation} QUBO problems were performed on a \jcmindex{DW}{D-Wave} D-Wave One and D-Wave Two machines without a solid demonstration of quantum speedup of annealer over classical algorithms \cite{Dwave3,Dwave4,Dwave5}. A better performance was shown for the last 2000-qubit \jcmindex{DW}{D-Wave} D-Wave machine released in 2017 on a newly proposed synthetic problem class in which the computational hardness is created through frustrated global interactions. The major limitations of \jcmindex{DW}{D-Wave} D-wave simulators is that each qubit can be connected to maximum of six other qubits which is the consequence of creating chips with Chimera structure. The next generation of \jcmindex{DW}{D-Wave} D-Wave quantum computer is expected to be announced in 2019 with a different architecture which would allow for 15 connections per each node. Together with reverse annealing and virtual graphs features a significant performance improvement could be possibly demonstrated.

\subsection{Complex \jcmindex{L}{laser}laser networks}
\label{subsec:2}
A new generation of complex \jcmindex{L}{laser}lasers such as degenerate cavity \jcmindex{L}{laser}lasers, multimode fibre amplifiers, large-aperture VCSEL, random \jcmindex{L}{laser}lasers have many advantages in comparison with the relatively simple  traditional
\jcmindex{L}{laser}laser resonators in terms of their computing properties \cite{controlcoh}.  They have a large
number of spatial degrees of freedom,   their \jcmindex{N}{nonlinear} nonlinear 
interactions within the gain material can be controlled by adjusting the spatial structures of lasing modes, 
the spatial coherence of emission can be tuned over
a wide range, and the output beams may have  arbitrary profiles. These properties allow the complex \jcmindex{L}{laser}lasers to be 
 used for reservoir computing \cite{photonicRC}  or for solving hard computational
problems. 

In \jcmindex{L}{laser}laser networks the  coupling can be engineered  by mutual light injection from one \jcmindex{L}{laser}laser to another. This introduces losses that depend on the relative phases between the \jcmindex{L}{laser}lasers. Such dissipative coupling drives the system to a phase locking and therefore to a steady state solution of \jcmindex{QCO}{quadratic continuous optimisation}  QCO (\ref{xy}), i.e. to the minimum of the \jcmindex{XY}{XY Hamiltonian} XY Hamiltonian \cite{lasers,lasers2,lasers3}. Degenerate cavity \jcmindex{L}{laser}lasers are particularly useful  as solvers as all their transverse modes have nearly identical quality factor. This implies that   a large number of transverse  modes  lase simultaneously since they all have  similar lasing thresholds \cite{controlcoh}.

The evolution of the $N$ single transverse and longitudinal modes class-B
\jcmindex{L}{laser}lasers can be described by the rate equations  \cite{ratelasers,nirprl2017}  on the amplitude $A_i$, phase $\theta_i$, and gain $G_i$ of the $i$-th \jcmindex{L}{laser}laser
\begin{eqnarray}
\frac{d A_i}{d t} &=&(G_i-\alpha_i)\frac{A_i}{\tau_p}+\sum_{j}J_{ij}\frac{A_j}{\tau_p}\cos(\theta_i-\theta_j),\label{l1}\\
\frac{d \theta_i}{d t} &=&\Omega_i-\sum_{j}J_{ij}\frac{A_j}{\tau_p A_i}\sin(\theta_i-\theta_j),\label{l2}\\
\frac{d G_i}{d t} &=&\frac{1}{\tau_c}[P_i-G_i(1+|A_i|^2)],
\label{l3}
\end{eqnarray}
where $P_i, \alpha_i, \Omega_i$ represent the pump strength, loss, frequency detuning of \jcmindex{L}{laser}laser $i$, respectively, whereas $\tau_p$ and $\tau_c$ denote the cavity round trip time and   the carrier
lifetime, respectively. The coupling strengths between $i$-th and $j$-th \jcmindex{L}{laser}lasers are represented by $J_{ij}$. 
If the amplitudes  of all \jcmindex{L}{laser}lasers are equal, Eq.~(\ref{l2}) 
reduces to the system of coupled
phase oscillators
\begin{equation}
\frac{d \theta_i}{d t} =\Omega_i-\frac{1}{\tau_p}\sum_{j}J_{ij}\sin(\theta_i-\theta_j).
\label{kuramoto}
\end{equation}
Equation (\ref{kuramoto}) is a celebrated Kuramoto model of identical oscillators which is widely used to describe the emergence of coherent  behaviour in complex systems \cite{kuramoto,kuramoto2}.   By LaSalle Invariance Principle  \cite{khalil} every trajectory of the Kuramoto model converges  to a minimum of the \jcmindex{XY}{XY Hamiltonian} XY Hamiltonian.

It was shown that the probability of finding the \jcmindex{GM}{global minimum} global minimum of the \jcmindex{XY}{XY Hamiltonian}  XY Hamiltonian agrees between experimental realisations of the \jcmindex{L}{laser}laser array and numerical simulations of Eqs.~(\ref{l1}-\ref{l3}). However,   simulating the Kuramoto model of Eq.~(\ref{kuramoto})    on the same matrix of coupling strengths gives a much lower probability of finding the \jcmindex{GM}{global minimum}  global minimum. The conclusion was made that the amplitude dynamics described by Eq.~(\ref{l1}) provides
a mechanism to reach the \jcmindex{GM}{global minimum}  global minimum \cite{nirprl2017} by pumping from below. This suggested that the cavity \jcmindex{L}{laser}lasers can be used as an efficient physical simulator for finding the \jcmindex{GM}{global minimum}  global minimum of the \jcmindex{XY}{XY Hamiltonian}  XY Hamiltonian, and therefore, for solving phase retrieval problems. 

A digital degenerate cavity \jcmindex{L}{laser}laser has recently been shown to solve phase retrieval problems rapidly \cite{nir2018}. 
It is an all-optical system that uses \jcmindex{N}{nonlinear} nonlinear  lasing process to find a  solution that best satisfies the constraint on the Fourier magnitudes of the light  scattered  from an object. To make sure that the solution to the phase retrieval problem is found the compact support aperture is introduced inside the cavity that ensures that  different configurations of \jcmindex{L}{laser}laser phases compete to find the one with the  minimal losses. The system combines the advantages of  short round-trip times of the order of 20ns and high parallelism in selecting the winning mode.

\subsection{Coherent Ising Machine}
\label{subsec:3}

Network of coupled \jcmindex{OPO}{optical parametric oscillator} optical parametric oscillators (OPOs) is an alternative physical system for solving the \jcmindex{IH}{Ising Hamiltonian}  Ising problem (\cite{CIM1} and references therein). Each \jcmindex{OPO}{optical parametric oscillator} OPO is a \jcmindex{N}{nonlinear} nonlinear  oscillator with two possible phase states above the threshold that can be interpreted as binary spin states $\{-1,1\}$ with respect to the reference beam. 
The \jcmindex{OPO}{optical parametric oscillator} OPO is stimulated with pulses of light which are then loaded into a loop of optical fiber. Below threshold, pulses of low intensity have random phase fluctuations. Depeding on the enforced pulse interactions, the intensities are continuously modulated so that after multiple runs around the loop the final binary phases are formed for all \jcmindex{OPO}{optical parametric oscillator} OPOs at about the same time. Driving the system close to this near-threshold regime, the lowest loss configuration state can be found. This state corresponds to the optimal solution of the \jcmindex{IH}{Ising Hamiltonian} Ising Hamiltonian and, therefore, the \jcmindex{OPO}{optical parametric oscillator} OPO-based simulator is known as the \jcmindex{CIM}{Coherent Ising Machine}  coherent Ising Machine (CIM).

The currently most successful implementations of \jcmindex{CIM}{Coherent Ising Machine} CIMs have been realised using a fiber-based degenerate optical parametric oscillators (DOPOs) and a measurement based feedback coupling, in which a matrix-vector multiplication is performed on a field-programmable gate array (FPGA) embedded in the feedback loop. The computational performance of such scalable optical processor, that is bounded by the electronic feedback, was demonstrated for various large-scale \jcmindex{IH}{Ising Hamiltonian} Ising problems \cite{CIM1,CIM3,CIM4}, while a speedup over classical algorithms is an ongoing study \cite{CIM5,Dwave_vs_CIM}. The ability to implement arbitrary coupling connections betwen any two spins \cite{CIM1} was apparently the main reason to claim a better scalability of the \jcmindex{CIM}{Coherent Ising Machine} CIM than the quantum annealer, i.e. \jcmindex{DW}{D-wave} D-Wave machine \cite{CIM3}.

In a \jcmindex{CIM}{Coherent Ising Machine} Coherent Ising Machine each Ising spin corresponds to a DOPO that is described by a rate equation for the complex amplitude of the signal field $a_i$:
\begin{equation}
	\frac{d a_i}{dt} = p a_i^* - a_i - |a_i|^2 a_i + \sum_j J_{ij} a_j,
	\label{CIM_ai}
\end{equation}
where the dynamics is defined by a linear pump term $p$, normalised linear and \jcmindex{N}{nonlinear} nonlinear  losses, and mutual couplings $J_{ij}$. To experimentally realise these couplings, a portion of light is extracted from the cavity after each round trip. That light is then homodyned against a reference pulse to produce $a_i$ that is next supplied to FPGA where a feedback signal is computed for each pulse. Lastly, an optical modulator is applied to convert the signal back to light that can be used for the next round trip. The equations (\ref{CIM_ai}) are often reformulated in terms of the in-phase and quadrature components $a_i = c_i + i s_i$ giving the equations in real terms:
\begin{eqnarray}
	\frac{d c_i}{dt} &=& \bigg(p - 1 - (c_i^2 + s_i^2) \bigg) c_i + \sum_j J_{ij} c_j \\
	\frac{d s_i}{dt} &=& \bigg(- p - 1 - (c_i^2 + s_i^2) \bigg) s_i + \sum_j J_{ij} s_j.
\end{eqnarray}
The computational effectiveness  of these equations has been demonstrated \cite{CIM_eqs2015} by tackling small size Ising type problems of order up to 20. In a part devoted to polariton \jcmindex{C}{condensate}condensates we will show that for achieving the \jcmindex{GM}{global minimum}  global minimum the realisation of an individual pump variation $p_i$ for equalising all signal amplitudes $|a_i|$ is crucial.

Phase-stability for the whole length of the cavity is required which makes the DOPOs system highly susceptible to external perturbations that can affect performance \cite{CIM3}. Furthermore, the \jcmindex{N}{nonlinear} nonlinear  DOPO generation process demands powerful \jcmindex{L}{laser}laser systems and temperature-controlled \jcmindex{N}{nonlinear} nonlinear  materials, which result in large and complex optical setups. These issues lead to recent proposals of other physical platforms for implementing a CIM-like machine. A CIM based on opto-electronic oscillators with self-feedback was suggested to be more stable and cheaper based on solving \jcmindex{IH}{Ising Hamiltonian} Ising optimisation problems on regular and frustrated graphs with up to 100 spins and similar or better performance compared to the original DOPO-based CIM \cite{CIM6}. An analogue all-optical implementation of a \jcmindex{CIM}{Coherent Ising Machine} CIM based on a network of injection-locked multicore fiber \jcmindex{L}{laser}lasers \cite{CIM7} demonstrated a possibility to solve \jcmindex{IH}{Ising Hamiltonian} Ising Hamiltonians for up to thirteen nodes. The dynamics of a network of injection-locked \jcmindex{L}{laser}lasers was based on \jcmindex{N}{nonlinear} nonlinear  coupled photon rate equations and the couplings were implemented using spatial light modulators (SLMs). The couplings were reported to be dependent on the photon numbers that are not known beforehand, which can be a major obstacle on the way of solving a given \jcmindex{IH}{Ising Hamiltonian} Ising Hamiltonian with the proposed photonic CIM. To solve this issue, approaches similar to gain variation \cite{GainD_sciRep_2018,NJP_Kalinin2018} may be considered in the future. Another large-scale optical Ising machine based on the use of an SLM was experimentally demonstrated by using the binary phases in separated spatial points of the optical wave front of an amplitude-modulated \jcmindex{L}{laser}laser beam and realising configurations with thousands of spins with tunable all-to-all pairwise interactions \cite{CIM8}.

\subsection{Photon and Polariton networks}
\label{subsec:4}

Microcavity \jcmindex{P}{polariton} exciton-polaritons, or simply \jcmindex{P}{polariton}  polaritons, are quasi-particles that result from the
hybridisation of light confined inside semiconductor microcavities and bound electron hole pairs (excitons). The steady states in these \jcmindex{NE}{nonequilibrium} nonequilibrium systems are set by the balance between the pumping intensity, coming from the interconversion rate of the exciton's reservoir into \jcmindex{P}{polariton}  polaritons, and losses, happening due to the leakage of photons. \jcmindex{P}{polariton}  Polaritons are bosons and obey \jcmindex{BS}{Bose-Einstein}  Bose-Einstein statistics, They can form a condensed (coherent) state above a critical density \cite{Kasprzak2006}. Thus, \jcmindex{P}{polariton} polaritons offer a unique playground to explore \jcmindex{NE}{nonequilibrium}  nonequilibrium condensation and related effects in solids. The advantage for such explorations comes from the \jcmindex{P}{polariton} polariton's small effective mass that is 4-5 orders of magnitude smaller than the electron's mass. The design and choice of material allows one to control the \jcmindex{P}{polariton} polariton mass and to realise such solid state \jcmindex{NE}{nonequilibrium} nonequilibrium \jcmindex{C}{condensate}condensates not only at cryogenic temperatures but even at room temperature in organic structures. The weak coupling at high temperatures 
and high pumping intensities transitions continuously to strong coupling at lower temperatures
and lower pumping intensities. In the limit of a small gain, i.e. small losses, solid state \jcmindex{C}{condensate}condensates resemble equilibrium Bose-Einstein \jcmindex{C}{condensate}condensates (BECs) and in the regime of high gain, i.e. high losses, they approach the \jcmindex{L}{laser}lasers. This transition from the equilibrium BECs to normal \jcmindex{L}{laser}lasers was described with a unified approach via \jcmindex{P}{polariton} polariton \jcmindex{C}{condensate}condensates \cite{UniversalityBook}.

%
In another system, closely resembling the physics of \jcmindex{P}{polariton} polariton \jcmindex{C}{condensate}condensates,  macroscopic occupation of the lowest mode for a gas of photons confined in a dye-filled optical microcavity was recently shown \cite{photoncondensate,photoncondensate2,photoncondensate3,photoncondensate4}. The rapid thermalization of rovibrational modes of the dye molecules by their collisions with the solvent and phonon dressing of the absorption and emission by the dye molecules leads to the thermal equilibrium distribution of photons and concomitant  accumulation of low-energy photons. Such systems resemble microlasers \cite{microlasers}, but unlike microlasers exhibit  a sharp threshold which occurs far below inversion.

To realise the lattices of \jcmindex{P}{polariton} polariton or photon \jcmindex{C}{condensate}condensates many techniques have been proposed and realised in experiments. \jcmindex{P}{polariton} Polariton lattices can be optically engineered by  injecting \jcmindex{P}{polariton}  polaritons  in specific areas of the sample using a spatial light modulator \cite{wertz,manni,pendulum,baumbergGeometrical,BerloffNatMat2017}. A variety of potential landscapes to confine \jcmindex{P}{polariton} polariton or photons have also been engineered  \cite{chneiderRev, amo,klaers17}. The rate equations describing the evolution of \jcmindex{GD}{gain-dissipative} gain-dissipative  \jcmindex{C}{condensate}condensates in a lattice were derived from the space and time resolved mean-field equations \cite{NJP_Kalinin2018, OurLattices} and take a form of the Stuart-Landau equations
\begin{equation}
\dot{\Psi_i}=-i U |\Psi_i|^2\Psi_i  + (\gamma_i- |\Psi_i|^2) \Psi_i + \sum_{j\ne i} {\cal C}_{ij}\Psi_j,\label{ee0}
\end{equation}
where $\Psi_i=\sqrt{\rho_i} \exp[i \theta_i]$ is  the complex amplitude of the $i-$th \jcmindex{C}{condensate}condensate, $U$ is the strength of self-interactions between the quasi-particles, $\gamma_i$ is the effective injection rate (the difference between the pumping of the quasi-particles into the system and linear losses). The coupling strength ${\cal C}_{ij}=J_{ij} + i G_{ij}$ is generally a complex number and consists of the Heisenberg coupling $J_{ij}$  mediated by the injection reservoir and the Josephson part $G_{ij}$ that comes from exchange interactions between the \jcmindex{C}{condensate}condensates. The system described by Eq. (\ref{ee0}) reaches the fixed point when $J_{ij}\gg G_{ij}$ and the pumping feedback is introduced in the system \cite{NJP_Kalinin2018}. The feedback on the pumping intensity ensures that all the occupations are the same at the fixed point, by adjusting the pumping if the occupation exceeds the set threshold  value $|\Psi_i|^2 = \rho_{\rm th}$. The total injection of the particles in the system of $N$ \jcmindex{C}{condensate}condensates at the fixed point is given by
\begin{equation}
\sum_{i=1}^N \gamma_i=N \rho_{\rm th}- \sum_{i=1}^N\sum_{j<i}^N J_{ij} \cos(\theta_i-\theta_j).
\label{cond}
\end{equation}
Choosing the lowest possible total particle injection $\sum \gamma_i$ that leads to the occupation $\rho_{\rm th}$ for each \jcmindex{C}{condensate}condensate guarantees that the minimum of the \jcmindex{XY}{XY Hamiltonian} XY Hamiltonian is reached. In order to find the true \jcmindex{GM}{global minimum}  global minimum the system has to slowly be brought to the condensation threshold while spending enough time in its neighbourhood to span various phase configurations driven by the system noise (classical and quantum fluctuations). When the system reaches a phase configuration in the vicinity of the minimum of the \jcmindex{XY}{XY Hamiltonian}  XY Hamiltonian it quickly converges to it by the gradient decent given by the imaginary part of Eq. (\ref{ee0}):
\begin{equation}
\dot{\theta_i}=-U \rho_{\rm th}- \sum_{j \ne i}^N J_{ij} \sin(\theta_i-\theta_j). \label{kuramoto}
\end{equation}
 This idea has been theoretically justified \cite{NJP_Kalinin2018} and experimentally realised for simple \jcmindex{PG}{polariton graphs} polariton graphs \cite{BerloffNatMat2017}.  It was also proposed how to extend the scheme to discrete \jcmindex{O}{optimisation} optimisation problems such as \jcmindex{QUBO}{quadratic unconstrained binary optimisation} QUBO (minimising the \jcmindex{IH}{Ising Hamiltonian} Ising Hamiltonian) or $n$-states \jcmindex{PH}{Potts Hamiltonian} Potts Hamiltonians \cite{GD-resonant}. When the resonant excitation is combined with a non-resonant one, the spins are forced to take the discrete values aligning with the directions set by the resonant excitation. If $n :1 $ resonant drive is added to the system, the dynamics of the coherent centres obeys
\begin{equation}
\dot{\Psi_i}=-i U |\Psi_i|^2\Psi_i  + (\gamma_i- |\Psi_i|^2) \Psi_i + \sum_{j\ne i} {J}_{ij}\Psi_j + h(t) \Psi_i^{*(n-1)},\label{res}
\end{equation}
where $h(t)$ is an increasing function that reaches some value $H>\max_i\sum_j |J_{ij}|$ at the threshold density $\rho_{th}$. The adjustment of injection rates leads to the equal-density fixed point and Eq.~(\ref{cond}) becomes
\begin{equation}
\sum_{i=1}^N \gamma_i=N \rho_{\rm th}- \sum_{i=1}^N\sum_{j<i}^N J_{ij} \cos(\theta_i-\theta_j)-H \rho_{\rm th}^{n/2-1}\cos(n\theta_i).
\label{cond2}
\end{equation} 
At $n=2$, the last term on the right-hand side provides the penalty to phases deviating from $0$ or $\pi$ reducing the \jcmindex{O}{optimisation} optimisation problem to \jcmindex{QUBO}{quadratic unconstrained binary optimisation} QUBO. For $n>2$, the $n$-state Potts Hamiltonian is minimised. The minimisation of HOBO may be achieved when the system operates much above the threshold and higher order terms can not be neglected \cite{stroev19}. 

If the time evolution of the reservoir of noncondensed particles is slow, the system of $N$ interacting coherent centres is better described by the 
following equations \cite{OurLattices}: 
\begin{eqnarray}
 \dot{\Psi_i}&=& -i U |\Psi_i|^2\Psi_i  +(R_i- \gamma_c) \Psi_i + \sum_{j\ne i} { J}_{ij}\Psi_j,\label{ee1}\\
 \dot{R_i}  &=&\Gamma_i- \gamma_R R_i -R_i|\Psi_i|^2,\label{ee2}
 \end{eqnarray}  
 where  $R_i$  is the occupation of the $i-$th  reservoir, $\Gamma_i, \gamma_R$ and $\gamma_c$ characterize the rate of particle injection into the reservoir and the linear losses of the reservoir and \jcmindex{C}{condensate}condensate, respectively.    If one replaces  $\Psi_i$  by the electric field and $R_i$ by the population inversion of the $i-$th \jcmindex{L}{laser}laser, the result is a form of the \jcmindex{LK}{Lang-Kobayashi} Lang-Kobayashi equations normally derived to describe the dynamical behavior of coupled \jcmindex{L}{laser}lasers from  Lamb's semiclassical \jcmindex{L}{laser}laser theory  \cite{lang80,acebron}.  The total injection of the particles in the system of $N$ \jcmindex{C}{condensate}condensates at the fixed point is given by
\begin{equation}
\sum_{i=1}^N \Gamma_i=(\gamma_R+\rho_{\rm th})[N \gamma_c- \sum_{i=1}^N\sum_{j<i}^N J_{ij} \cos(\theta_i-\theta_j)].
\end{equation}
Similar to Eq.~(\ref{cond}), if the total injection into the system is minimal, the phases of coherent centres minimise the \jcmindex{XY}{XY Hamiltonian} XY  Hamiltonian. 

Next we discuss the current and future role of physical systems mentioned above in a rapidly developed machine learning applications.

\section{Analogue Physical Systems for Recurrent Neural Networks and Reservoir Computing}
\label{sec:3}

The artificial neuron (perceptron) is a weighted decision-making procedure that takes binary input vector and produces a single binary output \cite{perceptron1962}. For a stable learning process, a small change in a weight (or bias) should cause only a small change in the output so that a few updates of parameters with a technique such as backpropagation \cite{backprop1986} can produce a better output. Such condition doesn't hold for a perceptron network that is sensitive to small changes in parameters of any single neuron. 
Introducing a \jcmindex{N}{nonlinear} nonlinear  activation function helps to overcome this problem with common choices for the function being the sigmoid, tanh, and the rectified linear unit.  With an addition of intermediate layers of neurons, hidden layers, the architecture and the training of the network can become nontrivial. The final design  usually is a trade off between the number of hidden layers against the time required to train the network. In fact, even optimising the design for single hidden-layer networks is an \jcmindex{NPH}{NP-hard} $\mathbb{NP}$-hard problem \cite{NN_npcomplete1989}. The outputs in each layer can be calculated given the outputs from the lower layers, so the information is always transmitted forward. 

Such feedforward neural networks are powerful learning models that have shown state-of-the-art results on a variety of machine learning applications \cite{ML_appls1} though they were still limited to the tasks with independent training and test data points. Addressing the data with time and space relationships (e.g. video frames and audio snippets) requires backward connection between layers of neurons, and so better modelled with \jcmindex{RNN}{recurrent neural networks} recurrent neural networks (RNNs).  

The capability of \jcmindex{RNN}{recurrent neural networks} RNNs with\jcmindex{N}{nonlinear} nonlinear  activations to perform nearly arbitrary computation was demonstrated with a simulation of a universal Turing machine \cite{RNN_Turing1991}. The two well-known \jcmindex{RNN}{recurrent neural networks} RNN architectures for sequence learning were introduced later: long short-term memory \cite{RNN_LSTM1997} and bidirectional \jcmindex{RNN}{recurrent neural networks} recurrent neural networks \cite{RNN_bidirect1997}. The former introduced the memory cell, a unit of computation that replaces traditional nodes in the hidden layer of a network, while the latter proposed that the output at any point in the sequence should be dependent on information from both the future and the past (comprehensive reviews of \jcmindex{RNN}{recurrent neural networks} RNNs can be found in \cite{RNN_review1,RNN_review2}). Following \cite{RNN_review2}, the input (target) sequence to a simple \jcmindex{RNN}{recurrent neural networks} RNN with one hidden layer can be denoted as a sequence of vectors ${\bf x}^{(t)}$ (${\bf y}^{(t)}$) for $t = \overline{1,T}$. The \jcmindex{RNN}{recurrent neural networks} RNN will then produce predicted vectors ${\bf y}^{(t)}$ at each time step:
\begin{eqnarray}
	{\bf h}(t) &=& f_{\rm activation}^{(1)} \bigg( W^{hx} {\bf x}(t) + W^{hh} {\bf h}(t-1) + {\bf b}_h \bigg) \\
	\widehat{{\bf y}}(t) &=& f_{\rm activation}^{(2)} \bigg( W^{yh} {\bf h}(t) + {\bf b}_y \bigg)
\end{eqnarray}
where $W^{hx}$ ($W^{yh}$) is the weight matrix between the input (output) and the hidden layer, $W^{hh}$ is the matrix of recurrent weights connecting the hidden layer with itself at adjacent time steps, ${\bf b}_h$ and ${\bf b}_y$ are bias vectors, and $f^{(i)}_{\rm activation}$ are \jcmindex{N}{nonlinear} nonlinear  activation functions. In such RNN, nodes with recurrent edges, i.e. edges that connect adjacent time steps, receive input from both the current state ${\bf x}(t)$ and from the previous state via hidden node values ${\bf h}(t-1)$. Given the hidden node values ${\bf h}(t)$, the output values $\widehat{{\bf y}}(t)$ are calculated which can be affected by the input ${\bf x}(t-1)$. Such cyclic network can still be trained across many time steps using backpropagation through time \cite{backprop_time} since \jcmindex{RNN}{recurrent neural networks} RNNs can be interpreted as a deep network with one layer per time step and shared weights across time steps.

\jcmindex{RNN}{recurrent neural networks} RNNs constitute a natural approach to numerous problems ranging from handwriting generation \cite{Graves2009,Graves2013} to character prediction \cite{Sutskever2011} to machine translation \cite{MachineTranslation2013,ML_appls2}. Such successful results and computational hardness of learning process make recurrent networks a great candidate for simulations with a physically based platform. Such physical systems can be much more efficient than any of known classical hardware implementations due to the inherent physical properties such as quantum or classical parallellism and neural network-type architecture.  A recent mapping of dynamics of acoustic and optical waves to \jcmindex{RNN}{recurrent neural networks} RNNs was suggested \cite{WavesToRNNs}.

Other computational frameworks for data processing such as \jcmindex{RC}{reservoir computing} reservoir computing (RC) (originally referred to as echo state networks \cite{ESN_2002} or liquid state machines \cite{LSM_2002}) can benefit from  an implementation with \jcmindex{APP}{analogue physical platforms} analogue physical platforms. A particular useful property is that \jcmindex{RC}{reservoir computing} RC systems demand lesser   precision of  individual control of all forward and backward neuron connections. 
Various \jcmindex{N}{nonlinear} nonlinear  dynamical systems, including 
electronic \cite{electronic_RC1,electronic_RC2,electronic_RC3}, 
photonic \cite{Photonic_RC1,Photonic_RC2}, 
spintronic \cite{Spintronic_RC1,Spintronic_RC2,Spintronic_RC3}, 
mechanical \cite{Mechanical_RC1}, 
and biological \cite{Biological_RC1} systems, 
have been recently employed as potential reservoirs for \jcmindex{RC}{reservoir computing} RC (see \cite{Review_physicalRC2018} and references therein). 

\jcmindex{RC}{reservoir computing} RC methods have been successfully applied to many practical problems involving real data, with focus on machine learning applications. 
The role of the reservoir (physical system) in \jcmindex{RC}{reservoir computing} RC is to \jcmindex{N}{nonlinearly} nonlinearly map sequential inputs into a higher-dimensional space so that features can then be extracted from its output with a simple learning algorithm. Therefore, such reservoirs become attractive for an experimental implementation in many physical systems with a motivation of realising fast information processing devices with low learning cost. Networks of \jcmindex{P}{polariton}  polaritons or  lattices of atomic \jcmindex{C}{condensate}condensates, discussed above, can serve as interacting \jcmindex{N}{nonlinear} nonlinear elements for an efficient network-type \jcmindex{RC}{reservoir computing} RC system with a possible approach suggested recently \cite{liewRC}.  
%

\section{System-inspired algorithms}
\label{sec:4}
The discovered principles of operation of the \jcmindex{APS}{analogue physical systems} analogue physical systems for finding \jcmindex{OS}{optimal solution} optimal solutions lead to the opportunity of formulating new \jcmindex{O}{optimisation} optimisation algorithms to be realised on specialised but classical computing architectures: FPGAs, GPAs, etc.

The principle of operation of the \jcmindex{CIM}{Coherent Ising Machine} Coherent Ising Machine was implemented as the network of \jcmindex{N}{nonlinear} nonlinear  oscillators 
described by  simplified equations \cite{CIM_eqs2017}:
\begin{equation}
	\frac{dx_j}{dt} = - \frac{\partial V}{\partial x_j} \quad {\rm with} \quad V = \sum_j V_b(x_j) + \epsilon V_{H}(x_j),
\end{equation}
where $x_j$ are $N$ analogue variables, $V_b(x_j) = -0.5 \alpha x_j^2 + 0.25 x_j^4$ is the paradigmatic bistable potential, $\alpha = -1 + p$ is the bifurcation parameter given by the normalized decay rate and linear gain $p$ for the signal field, $\epsilon \ll 1$ is a positive coefficient, $V_{H}(x_j) = - \sum_i J_{ij} x_i x_j$ is the analogue of the \jcmindex{HN}{Hopfield network} Hopfield network. The addition of the amplitude variation to these equations \cite{CIM_eqs2019}, i.e. $\epsilon \rightarrow\epsilon e_j$, resulted in 
\begin{equation}
	\frac{d e_j}{dt} = \beta (\rho_{\rm th} - x_j^2 ) e_j,
\end{equation}
where $e_i$ is a positive error variable and $\beta$ is a positive rate of change of error variables. The initialisation of such additional control of the target amplitude $\rho_{\rm th}$ of all analogue variables $x_j$ allowed to numerically optimise the medium-scale \jcmindex{IH}{Ising Hamiltonian} Ising type problems ($\sim1000$ spins).
A similar approach has been recently realised using FPGAs for a network of Duffing oscillators \cite{toshiba}.

Another example of a \jcmindex{GD}{gain-dissipative} gain-dissipative algorithm was inspired by the operation of \jcmindex{PN}{polariton networks} polariton networks \cite{GainD_sciRep_2018}. By gradually increasing the pumping strength $\gamma_i$ for $i-$th oscillator while inducing enough noise to span large volume of high dimensional space of the problem one can explore the low energy part of the spin Hamiltonian. This can be achieved by numerical integration of complex fields $\Psi_i$
\begin{eqnarray}
\dot{\Psi_i}&= &(\gamma_i- |\Psi_i|^2) \Psi_i + \sum_{j\ne i} {J}_{ij}\Psi_j + h(t) \Psi_i^{*(n-1)},\label{res2}\\
\dot{\gamma_i}&=&\epsilon(\rho_{\rm th}-\rho_i),\label{gamma}
\end{eqnarray}
with the meaning of parameters explained above. The performance of this algorithm was demonstrated on the medium scale \jcmindex{IH}{Ising Hamiltonian} Ising and \jcmindex{XY}{XY Hamiltonian}  XY models  \cite{GainD_sciRep_2018}.


\section{Conclusions and Future Challenges}

What promise does \jcmindex{UC}{unconventional computing} unconventional computing platform hold? Would it be able to find a better solution in a fixed time? Or find a solution for a fixed precision faster? Or solve more complex problems at fixed and limited cost? Would it appear as an accelerator in \jcmindex{NN}{neural networks}  neural networks, \jcmindex{ML}{machine learning} machine learning and artificial intelligence platforms? Would it be able to solve a full range of different problems or perform a single computationally intensive task or operation as a part of a hybrid platform?
The answer "yes" to any of these challenges could imply high technological gains from predicting   and developing new materials and designs to creating fully automated AI-controlled systems.  

In the short term one may anticipate appearance of a plethora  of disparate \jcmindex{UC}{unconventional computing} unconventional computing systems designed to execute specific algorithms or solving a particular practically relevant task. These will be more energy-efficient and will demonstrate a superior performance over classical hardware architectures. These systems could be further orchestrated together to perform larger tasks. The role of orchestra conductor could be devoted to traditional computing resources including CPUs, GPUs, FPGAs, and others. Such symbiosis of classical and \jcmindex{UH}{unconventional harware} unconventional hardware together with the development of \jcmindex{SIA}{system-inspired algorithms} system-inspired optimisation algorithms will form an ultimate  hybrid  computing platform that may allow for continued scaling beyond the physical limits of Moore's law. Many recent proposals exploit this idea, including, for instance,  a photonic accelerator for neural networks with weights and inputs encoded in optical signals allowing for the neural network to be reprogrammed and trained on the fly at high speed \cite{hamerly2019large} and  the recurrent Ising machine in a photonic integrated circuit  \cite{prabhu2019recurrent}.

To characterise the advantages of \jcmindex{UC}{unconventional computing} unconventional computing  platforms we can apply  the  \jcmindex{QS}{quantum supremacy} "quantum supremacy" test formulated by John Preskill in 2012 to characterise superior performance of a \jcmindex{QS}{quantum simulator} quantum simulator over any existing classical computing machines \cite{preskill2012quantum}.   A particular milestone of \jcmindex{QS}{quantum supremacy} quantum supremacy has been recently demonstrated with the Google's Sycamore processor  \cite{arute2019quantum} based on 53 programmable superconducting qubits for a certain application, namely random circuit sampling. Their claimed speedup of 200$s$ against 10000 years was immediately scaled down to 200$s$ compared to 2 days by using the absolute state-of-the-art classical supercomputer ``Summit" at Oak Ridge National Lab \cite{pednault2019leveraging}. Nevertheless, this is still a remarkable, at least three orders of magnitude acceleration provided by \jcmindex{QS}{quantum simulations} quantum simulations in comparison with classical conventional computing. This achievement is bound to ignite a \jcmindex{QS}{quantum supremacy} quantum supremacy race, as already shown by the subsequent boson sampling experiment with tens  of photons covering the similar sized Hilbert space as with 48 qubits \cite{wang2019boson}. In upcoming years, we envision many more systems to actively participate in this race and demonstrate their superior suitability for solving particular classes of problems.

The \jcmindex{UC}{unconventional computing} unconventional computing systems we described in this Chapter are not quantum computers in the traditional meaning of this term. Although quantum effects contribute to the system operation (e.g. Bose-Einstein condensation is a quantum process that obeys quantum statistics), it is not clear if these system offer any quantum speed-up during the search for a  solution, e.g. via entanglement and superposition of states. Although, unlike quantum computers these systems  have a crucial ingredient that drives their operation: \jcmindex{N}{nonlinearity} nonlinearity! \jcmindex{N}{nonlinearity} Nonlinearity leads to the emergence of coherence within each "bit" of the \jcmindex{UC}{unconventional computing} unconventional computers we discussed: a \jcmindex{L}{laser} laser, a \jcmindex{C}{condensate}condensate, an optical parametric oscillator. \jcmindex{N}{nonlinearity} Nonlinearity drives the gain saturation, mode locking, and coupling. 

The physical systems that we described aim at finding the \jcmindex{GM}{global minimum}  global minimum of hard \jcmindex{O}{optimisation} optimisation problems. All these systems have advantages and limitations. They vary in \jcmindex{N}{nonlinearity} nonlinearity of the underlying modes of operation,  scalability, ability to engineer the required couplings, flexibility of turning the interactions, precision of read-out, factors facilitating the approach the \jcmindex{GM}{global minimum}  global rather than local minimum. These issues have to be addressed from the experimental point of view in all newly proposed physical platforms. 

All of the considered systems in this Chapter have some parts of their operation that promise increased performance over the classical computations. Combined with \jcmindex{SIA}{system-inspired algorithms} system-inspired computational algorithms these systems may indeed one day revolutionise our computing.

\section{Abbreviations}

\begin{description}[CABR]
\item[QP]{Quadratic Programming}
\item[DNLS]{Discrete Nonlinear Schr\"odinger Equation}
\item[QUBO]{Quadratic Unconstrained Binary Optimisation}
\item[QCO]{Quadratic Continuous Optimisation}
\item[HOBO]{Higher Order Binary Optimisation}
\item[CIM]{Coherent Ising Machine}
\item[OPO]{Optical Parametric Oscillator}
\item[DOPO]{Degenerate Optical Parametric Oscillator}
\item[FPGA]{Field-Programmable Gate Array}
\item[SDP]{Semidefinite Programming}
\item[MaxCut]{Maximum Cut}
\item[GPE]{Gross-Pitaevskii equation}
\item[RC]{Reservoir Computing}
\item[RNN]{Recurrent neural networks}

\end{description}

\bibliography{refs}
\bibliographystyle{plain}

\backmatter

\printindex

\end{document}